# Approximately Efficient Cost-Sharing Mechanisms

Tim Roughgarden*    Mukund Sundararajan[†]

August 22, 2018

**Abstract**

We study cost-sharing mechanisms for several fundamental NP-hard combinatorial optimization problems. A cost-sharing mechanism is a protocol that, given bids for a service, determines which bidders to serve and what prices to charge. The mechanism incurs a subset-dependent cost that is implicitly defined by an instance of a combinatorial optimization problem. Three desirable but mutually incompatible properties of a cost-sharing mechanism are: incentive-compatibility, meaning that players are motivated to bid their true private value for receiving the service; budget-balance, meaning that the mechanism recovers its incurred cost with the prices charged; and efficiency, meaning that the cost incurred and the valuations of the players served are traded off in an optimal way.

Our work is motivated by the following fundamental question: for which cost functions, and in what senses, are incentive-compatible mechanisms with good approximate budget-balance and efficiency possible? We make three different types of contributions to this question.

- We identify several new classes of combinatorial cost functions that admit incentive-compatible mechanisms achieving both a constant-factor approximation of budget-balance and a polylogarithmic approximation of the social cost formulation of efficiency. In particular, we exhibit such mechanisms for the classes of facility location and single-sink rent-or-buy cost functions, with approximate efficiency $O(\log k)$ and $O(\log^2 k)$, respectively. (Here $k$ denotes the number of players served in an efficient solution.) The mechanisms belong to the class of Moulin mechanisms, and therefore satisfy a strong version of incentive-compatibility known as groupstrategyproofness.

- We prove a new, optimal lower bound of $\Omega(\log^2 k)$ on the approximate efficiency of every $O(1)$-budget-balanced Moulin mechanism for Steiner tree or SSRoB cost functions. This lower bound exposes a latent approximation hierarchy among different cost-sharing problems.

- We show that weakening the definition of incentive-compatibility to strategyproofness can permit exponentially more efficient approximately budget-balanced mechanisms, in particular for set cover cost-sharing problems.

*Department of Computer Science, Stanford University, 462 Gates Building, 353 Serra Mall, Stanford, CA 94305. Supported in part by ONR grant N00014-04-1-0725, DARPA grant W911NF-05-1-0224, and an NSF CAREER Award. Email: tim@cs.stanford.edu.

[†]Department of Computer Science, Stanford University, 470 Gates Building, 353 Serra Mall, Stanford, CA 94305. Supported in part by OSD/ONR CIP/SW URI "Software Quality and Infrastructure Protection for Diffuse Computing" through ONR Grant N00014-01-1-0795 and by OSD/ONR CIP/SW URI "Trustworthy Infrastructure, Mechanisms, and Experimentation for Diffuse Computing" through ONR Grant N00014-04-1-0725. Email: mukunds@cs.stanford.edu.

# 1 Introduction

**Cost-Sharing Mechanisms.** We study cost-sharing mechanisms for several fundamental NP-hard combinatorial optimization problems. A *cost-sharing mechanism* is a protocol that collects bids for a service from potential users (players), chooses the subset of players to receive the service and a feasible way of servicing them, and determines prices to charge the chosen players. The mechanism incurs a subset-dependent cost $C(S)$ that is given by a known cost function $C$. In this paper, we consider cost functions that are defined implicitly by an instance of a combinatorial optimization problem.

Designing a cost-sharing mechanism requires picking and choosing between several desirable properties that have long been known to be mutually incompatible. The most commonly considered such properties are: (1) *incentive-compatibility*, meaning that players are motivated to bid their true private value $v_i$ for receiving the service; *budget-balance*, meaning that the mechanism recovers its incurred cost with the prices charged; and (3) *efficiency*, meaning that the cost incurred and the valuations of the players served are traded off in an optimal way.

The properties (1)–(3) cannot be simultaneously achieved, even for very simple cost functions and weak notions of incentive compatibility [5, 22]. This impossibility result motivated two distinct approaches to designing cost-sharing mechanisms. The first approach ignores budget-balance and seeks incentive-compatible and efficient cost-sharing mechanisms. The *VCG mechanism* (see e.g. [18, 20]) is the most flexible and well-known mechanism of this type. These mechanisms are typically *strategyproof (SP)*, in the sense that the utility of each player—its value $v_i$ minus the price it is charged, or 0 if it does not receive the service—is maximized by bidding truthfully for every fixed set of bids by the other players. They are typically not approximately budget-balanced for any reasonable approximation factor (see e.g. [3]).

The second design approach to cost-sharing mechanisms is to insist on incentive-compatibility and budget-balance, while regarding efficiency as a secondary objective. Moulin [19] introduced a class of mechanisms of this type, which have good budget-balance and are incentive compatible. In fact, Moulin mechanisms are *groupstrategyproof (GSP)*, meaning that an analogue of the SP condition holds even for coordinated bidding by coalitions of players. Researchers have developed approximately budget-balanced Moulin mechanisms for a number of different combinatorial optimization problems, including fixed-tree multicast [1, 3, 4]; the more general submodular cost-sharing problem [19, 20]; Steiner tree [11, 12, 14]; Steiner forest [15, 16]; facility location [17, 21]; rent-or-buy network design [9, 21], and various covering problems [2, 10]. Almost all known GSP cost-sharing mechanisms are Moulin mechanisms (see [10]).

**Approximately Efficient Cost-Sharing Mechanisms.** By design, budget-balance takes explicit precedence over efficiency in Moulin mechanisms. Indeed, nearly all previous papers that design Moulin mechanisms do not address the efficiency of the proposed mechanisms. Nevertheless, very recent work [23] shows that many Moulin mechanisms possess *approximately optimal* efficiency. The first evidence that Moulin mechanisms can have good efficiency properties was provided earlier by Moulin and Shenker [20], who gave a bound on the worst-case additive efficiency loss of a particular Moulin mechanism for submodular cost functions.

Several definitions of approximate efficiency are possible, and the choice of definition is important for quantifying the inefficiency of Moulin mechanisms. The most common formulation of (exact) efficiency is to require that a cost-sharing mechanism choose a subset of players that maximizes the *social welfare*, where the social welfare $W(S)$ of a set $S$ is defined as $\sum_{i \in S} v_i - C(S)$. Feigenbaum et al. [3] showed that, even for extremely simple cost functions, for every pair of constants $\alpha, \beta \geq 1$, no $\beta$-budget balanced SP mechanism always recovers an $\alpha$ fraction of the maximum social welfare. (A mechanism is $\beta$-budget-balanced if the sum of prices charged is always at most the cost incurred and at least a $\beta$ fraction of this cost.)

An alternative formulation of exact efficiency is to choose a subset minimizing the *social cost*, where we define the social cost $\pi(S)$ of a set $S$ as the sum of the incurred service cost and the excluded valuations:

$C(S) + \sum_{i \notin S} v_i$. Since $\pi(S) = -W(S) + \sum_{i \in U} v_i$ for every set $S$, where $U$ denotes the set of all players, a subset maximizes the social welfare if and only if it minimizes the social cost. In previous work [23], the authors showed that well-known, approximately budget-balanced Moulin mechanisms for submodular cost-sharing [20] and Steiner tree cost-sharing problems [11] always achieve a polylogarithmic (multiplicative) approximation of the optimal social cost. If a cost-sharing mechanism is an $\alpha$-approximation algorithm with respect to the social cost objective, then we say that the mechanism is $\alpha$-*approximate*.

The results in [23] raise the possibility of obtaining incentive-compatible mechanisms that simultaneously approximate budget-balance and efficiency for a wide array of important cost-sharing problems. At the highest level, this paper is motivated by the following fundamental question:

> *for which cost functions, and in what senses, are incentive-compatible mechanisms with good approximate budget-balance and efficiency possible?*

More precisely, our work addresses the following issues.

(1) Which cost functions admit a Moulin mechanism that is $\alpha$-approximate for a reasonable (say, polylogarithmic) factor $\alpha$? A necessary condition for this is that the cost function admit a so-called cross-monotonic cost-sharing method with good budget-balance [19, 23]; many but not all combinatorial cost-sharing problems admit such methods [10]. Is this condition also sufficient, or are there further obstructions to the design of approximately efficient Moulin mechanisms?

(2) Do combinatorial cost-sharing problems exhibit a natural "approximation hierarchy"? Do the twin requirements of incentive-compatibility and good budget-balance force quantitatively different amounts of efficiency loss in different combinatorial optimization problems?

(3) Several natural cost-sharing problems do not even admit Moulin mechanisms with good budget-balance [10]. Does weakening the incentive-compatibility constraint (from GSP to SP, say) permit approximately efficient and budget-balanced mechanisms for such problems?

**Our Results.** We make three different types of contributions on approximately efficient cost-sharing mechanisms.

First, we identify several new classes of combinatorial cost functions that admit $O(\text{polylog}(k))$-approximate Moulin mechanisms, where $k$ is the number of players served in an optimal solution. Previously, only submodular cost functions and Steiner tree cost functions were known to admit such mechanisms [23].

Such mechanisms can only exist for cost functions that admit $O(\text{polylog}(k))$-budget-balanced cross-monotonic cost-sharing methods [19, 23]. Our results nearly exhaust the combinatorial optimization problems for which such methods are known, leaving only the class of Steiner forest cost-sharing problems [15] unresolved.

In particular, for facility location cost-sharing problems, we show that the Moulin mechanism based on the $O(1)$-budget-balanced cross-monotonic cost-sharing method of Pál and Tardos [21] is $O(\log k)$-approximate. Our proof uses, as a black box, a result in [23] that reduces upper bounding the approximate efficiency of a Moulin mechanism to the "summability" of its underlying cost-sharing method (see Definition 2.3). The essence of this result is therefore our (problem-specific) argument that the Pál-Tardos cost-sharing method is $O(\log k)$-summable. A simple example shows that every $O(1)$-budget-balanced Moulin mechanism for facility location problems is $\Omega(\log k)$-approximate.

We also prove a bound of $O(\log^2 k)$ on the summability of the $O(1)$-budget-balanced cross-monotonic cost-sharing method for single-sink rent-or-buy (SSRoB) cost functions due to Gupta, Srinivasan, and Tardos [9] and Leonardi and Schäfer [17]. The main result in [23] then implies that the corresponding Moulin mechanism is $O(\log^2 k)$-approximate.

Second, we prove a new lower bound that exposes a non-trivial, latent hierarchy among different cost-sharing problems. Specifically, we prove that every $O(1)$-budget-balanced Moulin mechanism for Steiner tree cost functions is $\Omega(\log^2 k)$-approximate. This lower bound trivially also applies to SSRoB cost functions. A lower bound of $\Omega(\log^2 k)$ was previously known only for specific Steiner tree Moulin mechanisms [11, 23].

This lower bound establishes a previously unobservable separation between submodular and facility location cost-sharing problems on the one hand, and Steiner tree and SSRoB cost-sharing problems on the other. All four of these classes of cost functions admit $O(1)$-budget-balanced, cross-monotonic cost-sharing methods. However, the first two problems admit $\Theta(\log k)$-approximate Moulin mechanisms, while the latter two only allow $\Theta(\log^2 k)$-approximate Moulin mechanisms.

All previous lower bounds on the efficiency of Moulin mechanisms arose as a consequence of either budget-balance lower bounds or, in the case of the combinatorial optimization problems considered in this paper, from a trivial example equivalent to a single-link instance of fixed-tree multicast [23]. This type of example cannot be used to prove a lower bound larger than the $k$th Harmonic number $\mathcal{H}_k = \Theta(\log k)$ on the approximate efficiency of a Moulin mechanism. We obtain the stronger bound of $\Omega(\log^2 k)$ by a significantly more intricate construction that is specific to Steiner tree cost functions.

Third, we show that weakening the definition of incentive-compatibility can permit exponentially more efficient mechanisms for basic cost-sharing problems. For set cover cost-sharing problems, results of Immorlica, Mahdian, and Mirrokni [10] imply that every Moulin mechanism is $\Omega(\sqrt{k})$-budget-balanced and hence $\Omega(\sqrt{k})$-approximate [23]. Recall that essentially all known GSP mechanisms are Moulin mechanisms (see [10]). On the other hand, we show that the set cover mechanism of Devanur, Mihail, and Vazirani [2], which is SP and $O(\log k)$-budget-balanced but not GSP, is $O(\log k)$-approximate. We also show a similar, if less dramatic, result for facility location problems: SP mechanisms can obtain approximation factors that are unachievable with Moulin mechanisms.

## 2 Preliminaries

**Cost-Sharing Mechanisms.** We consider a cost function $C$ that assigns a cost $C(S)$ to every subset $S$ of a universe $U$ of players. We will always assume that $C$ is nonnegative and nondecreasing (i.e., $S \subseteq T$ implies $C(S) \leq C(T)$). We sometimes refer to $C(S)$ as the *service cost*, to distinguish it from the social cost (defined below). We also assume that every player $i \in U$ has a private, nonnegative *valuation* $v_i$.

A *(direct revelation) mechanism* collects a nonnegative bid $b_i$ from each player $i \in U$, selects a set $S \subseteq U$ of players, and charges every player $i$ a price $p_i$. We interpret the valuation $v_i$ as player $i$'s willingness to pay to be included in the chosen set $S$, and a mechanism typically uses the (known) bid $b_i$ as a proxy for the (unknown) valuation $v_i$. We will only allow mechanisms that are "individually rational" in the sense that $p_i = 0$ for players $i \notin S$ and $p_i \leq b_i$ for players $i \in S$. We also require that all prices are nonnegative ("no positive transfers"). As is standard, we assume that players have *quasilinear* utilities, meaning that each player $i$ aims to maximize $u_i(S, p_i) = v_i x_i - p_i$, where $x_i = 1$ if $i \in S$ and $x_i = 0$ if $i \notin S$.

Our first definition states that a mechanism should be designed to motivate players to bid truthfully (i.e., to set $b_i = v_i$ for all $i$). We will use the strong and well-known such definition of a *groupstrategyproof (GSP)* mechanism, which formalizes the idea of a "collusion-resistant" mechanism. The definition states that every coordinated set of false bids by a coalition should decrease the utility of some player in the coalition (or should have no effect).

**Definition 2.1** A mechanism is *groupstrategyproof (GSP)* if the following property holds for every input $(U, C, v)$. Let $T \subseteq U$ be a coalition of players. Let $b$ and $b'$ be two bid vectors (indexed by $U$) for which $b_i = b'_i$ for all players $i \notin T$ outside the coalition. Assume that players of $T$ bid truthfully in $b$ ($b_i = v_i$ for all $i \in T$). Let $(S, p)$ and $(S', p')$ be the mechanism outcomes given bids $b$ and $b'$, respectively. If $u_i(S', p') > u_i(S, p)$ for some player $i \in T$, then $u_j(S', p') < u_j(S, p)$ for some other player $j \in T$.

A *strategyproof (SP)* mechanism is only required to satisfy the constraint in Definition 2.1 for coalitions $T$ that are singletons.

There is a well-known tension between the goals of budget-balance and efficiency in cost-sharing mechanisms. For a parameter $\beta \geq 1$, a mechanism is $\beta$-*budget balanced* if for every possible outcome $(S, p)$ of the mechanism, it recovers at least a $1/\beta$ fraction of the service cost, with the constraint that the total cost-recovery does not exceed the service cost: $C(S)/\beta \leq \sum_{i \in S} p_i \leq C(S)$.

We measure efficiency using the objective of *social cost minimization*. Given a universe $U$ of players with valuations $v$ and a cost function $C$, the social cost of a set $S \subseteq U$ is defined by $\pi(S) = C(S) + \sum_{i \notin S} v_i$. We say that a cost-sharing mechanism is *efficient* if, assuming truthful bids, it always outputs a set $S$ that minimizes the social cost. Equivalently, an efficient mechanism always maximizes the *social welfare* $W(S) = \sum_{i \in S} v_i - C(S)$. Finally, a cost-sharing mechanism is $\alpha$-*approximate* if, assuming truthful bids, it always outputs a set $S$ with social cost at most an $\alpha$ factor times the optimal social cost.

For example, the *VCG mechanism* is efficient, is not $\beta$-budget-balanced for any finite factor $\beta \geq 1$, and is SP but not GSP (see e.g. [18, 20]). Below we review a class of mechanisms that are GSP and, for many cost functions, both approximately budget-balanced and approximately efficient.

In this paper, we focus on cost functions that are defined implicitly as the optimal solution of an instance of a (NP-hard) combinatorial optimization problem. Besides selecting the set of players to be served, the mechanism must also decide *how* to service this set, ideally in polynomial time. For instance, for set cover cost functions (Section 5), the cost $C(S)$ of serving a set $S$ of players is defined as the minimum cost of a collection of sets that covers all of $S$. The (polynomial-time) mechanism that we consider uses a greedy algorithm to construct a feasible (not necessarily optimal) cover of the chosen players. In such cases, we write $C'(S)$ for the service cost incurred by the mechanism, reserving $C(S)$ for the cost of the optimal way of providing service to the set $S$. A mechanism is now defined to be $\alpha$-*approximate* only if it always chooses a set $S$ and incurs a service cost $C'(S)$ such that $C'(S) + \sum_{i \notin S} v_i$ is at most $\alpha$ times the optimal social cost $\min_{S^* \subseteq U}[C(S^*) + \sum_{i \notin S^*} v_i]$. Similarly, a mechanism is $\beta$-budget-balanced only if it always recovers at least a $\beta$ fraction of the incurred cost $C'(S)$ while recovering no more than the cost $C(S)$ of optimally serving the set $S$.

**Moulin Mechanisms and Cross-Monotonic Cost-Sharing Methods.** Next we review a class of cost-sharing mechanisms called *Moulin mechanisms*. Almost all known GSP cost-sharing mechanisms are Moulin mechanisms; see [10, 19, 20] for detailed discussions on the relationship between Moulin mechanisms and general GSP cost-sharing mechanisms.

Given a universe $U$ of players and a cost function $C$, a *cost-sharing method* $\chi$ assigns a non-negative *cost share* $\chi(i, S)$ for every subset $S \subseteq U$ of players and every player $i \in S$. (Note players no longer have valuations.) A cost-sharing method is $\beta$-*budget balanced for* $C$ for a parameter $\beta \geq 1$ if it always recovers at least $\beta$ fraction of the cost, without recovering more than the cost: $C(S)/\beta \leq \sum_{i \in S} \chi(i, S) \leq C(S)$. A cost-sharing method is *cross-monotonic* if adding players to the set $S$ only decreases the cost shares of players: for all $S \subseteq T \subseteq U$ and $i \in S$, $\chi(i, S) \geq \chi(i, T)$.

A cost-sharing method $\chi$ for $C$ gives rise the following *Moulin mechanism* $M_\chi$ for $C$. First, collect a bid $b_i$ for each player $i$. Initialize the set $S$ to all of $U$ and set a price $p_i$ equal to $\chi(i, S)$ for each player $i$. If $p_i \leq b_i$ for all $i \in S$, halt and output the set $S$ (and charge prices $p$). If $p_i > b_i$ for some player $i$, then remove an arbitrary such player from the set $S$ and iterate. Moulin [19] proved the following.

**Theorem 2.2 ([19])** *If $\chi$ is a cross-monotonic cost-sharing method for the cost function $C$, then the corresponding Moulin mechanism $M_\chi$ is groupstrategyproof. Moreover, if $\chi$ is $\beta$-budget-balanced, then so is $M_\chi$.*

A cost-sharing method $\chi$ is said to be *in the core* if every subset $S'$ of players pays at most the cost $C(S')$ of optimally serving them alone: for all sets $S \subseteq U$ and for all $S' \subseteq S$, $\sum_{i \in S'} \chi(i, S) \leq C(S')$. The core condition is a natural fairness condition and has been extensively studied. It also turns out to be useful in proving upper bounds on the approximate efficiency of a mechanism. It is easy to see that every approximately budget-balanced and cross-monotonic cost-sharing method is in the core. The non-Moulin mechanisms that we study in Section 5 will also charge prices that satisfy the core condition.

**Summability and Approximate Efficiency.** The approximate efficiency of Moulin mechanisms was previously studied in [23]. The main result in [23] shows that the approximate efficiency of a Moulin mechanism is completely controlled by the budget-balance and one additional parameter of the underlying cost-sharing method. We define this parameter next.

**Definition 2.3 ([23])** Let $C$ and $\chi$ be a cost function and a cost-sharing method, respectively, defined on a common universe $U$ of players. The method $\chi$ is $\alpha$-*summable for $C$* if

$$\sum_{\ell=1}^{|S|} \chi(i_\ell, S_\ell) \leq \alpha \cdot C(S) \qquad (1)$$

for every ordering $\sigma$ of $U$ and every set $S \subseteq U$, where $S_\ell$ and $i_\ell$ denote the set of the first $\ell$ players of $S$ and the $\ell$th player of $S$ (with respect to $\sigma$), respectively.

The main result in [23] is as follows.

**Theorem 2.4 ([23])** *Let $U$ be a universe of players and $C$ a nondecreasing cost function on $U$ with $C(\emptyset) = 0$. Let $M$ be a Moulin mechanism for $C$ with underlying cost-sharing method $\chi$. Let $\alpha \geq 0$ and $\beta \geq 1$ be the smallest numbers such that $\chi$ is $\alpha$-summable and $\beta$-budget-balanced. Then the mechanism $M$ is $(\alpha + \beta)$-approximate and no better than $\max\{\alpha, \beta\}$-approximate.*

Note that the upper and lower bounds in Theorem 2.4 differ by at most a factor of 2. In particular, an $O(1)$-budget-balanced Moulin mechanism is $\Theta(\alpha)$-approximate if and only if the underlying cost-sharing method is $\Theta(\alpha)$-summable. Strictly speaking, the lower bound of $\alpha$ in Theorem 2.4 requires that the cost-sharing method $\chi$ satisfy at least one of several possible weak technical conditions, such as the notion of "strong consumer sovereignty" studied (in somewhat different forms) in [10, 20]. Since we will only use this characterization in situations where these technical conditions hold, we defer further discussion of this issue to the full version.

## 3 Approximately Efficient Cost-Sharing Mechanisms for Facility Location and Single-Sink Rent-or-Buy

### 3.1 A $\Theta(\log k)$-Approximate Mechanism for Facility Location

In this subsection we consider facility location cost-sharing problems. The input is given by a set $U$ of demands (the players), a set $F$ of facilities, an opening cost $f_q$ for each facility $q \in F$, and a metric $c$ defined on $U \cup F$. The cost $C(S)$ of a subset $S \subseteq U$ of players is then defined as the cost of an optimal solution to the uncapacitated facility location problem induced by $S$. In other words,

$$C(S) = \min_{\emptyset \neq F^* \subseteq F} \left( \sum_{q \in F^*} f_q + \sum_{i \in S} \min_{q \in F^*} c(q, i) \right).$$

We seek a cross-monotonic, approximately budget-balanced cost-sharing method for such problems that also has good summability. We begin with a simple lower bound, similar to that given in [23] for submodular cost-sharing problems.

**Proposition 3.1** *For every $k \geq 1$, there is a $k$-player facility location cost function $C$ with the following property: for every $\beta \geq 1$ and every $\beta$-budget-balanced, cross-monotone cost-sharing method $\chi$ for $C$, $\chi$ is no better than $\mathcal{H}_k/\beta$-summable.*

The proof of Proposition 3.1, and all other proofs of this paper, are deferred to the Appendix.

Applying the characterization in Theorem 2.4 yields a lower bound on the approximate efficiency of all Moulin mechanisms for facility location cost-sharing problems.

**Corollary 3.2** *For every $k \geq 1$, there is a $k$-player facility location cost function $C$ with the following property: for every $\beta \geq 1$ and every $\beta$-budget-balanced Moulin mechanism $M$ for $C$, $M$ is no better than $\mathcal{H}_k/\beta$-approximate.*

Pál and Tardos [21] showed the every facility location cost function admits a 3-budget-balanced cross-monotonic cost-sharing method $\chi_{PT}$. (We review this method in the proof of Lemma 3.5 in the Appendix.) Our first main result shows that the corresponding Moulin mechanism matches the lower bound in Corollary 3.2, up to a constant factor.

**Theorem 3.3** *Let $C$ be a $k$-player facility location cost function and $\chi_{PT}$ the corresponding Pál-Tardos cost-sharing method. Then $\chi_{PT}$ is $\mathcal{H}_k$-summable for $C$.*

Applying Theorem 2.4 yields an efficiency guarantee for the corresponding Moulin mechanisms.

**Corollary 3.4** *Let $C$ be a $k$-player facility location cost function and $M_{PT}$ the Moulin mechanism based on the corresponding Pál-Tardos cost-sharing method. Then $M_{PT}$ is GSP, 3-budget-balanced, and $(\mathcal{H}_k + 3)$-approximate.*

Theorem 3.3 will follow immediately from two lemmas. The first, and arguably more surprising, lemma states that single-facility instances supply worst-case examples for the summability of the PT cost-sharing scheme.

**Lemma 3.5** *For every $k \geq 1$, the summability of PT cost-sharing methods for $k$-player facility location cost functions is maximized by the cost functions that correspond to single-facility instances.*

Lemma 3.5 is based on a monotonicity property that we prove for the PT cost-sharing method: increasing the distance between a demand and a facility can only increase cost shares. While intuitive and easy to prove, analogous monotonicity properties fail for other cross-monotonic cost-sharing methods, such as the JV Steiner tree method [11].

Informally, this monotonicity property allows us to argue that in worst-case facility location instances, players are partitioned into non-interacting groups, each clustered around one facility. We complete the proof by arguing that the summability of the PT cost-sharing method for one of these single-facility clusters in at least that in the original facility location instance.

Our second lemma bounds the summability of PT cost-sharing methods in single-facility instances.

**Lemma 3.6** *Let $C$ be a $k$-player facility location cost function corresponding to a single-facility instance. If $\chi_{PT}$ is the corresponding PT cost-sharing method, then $\chi_{PT}$ is $\mathcal{H}_k$-summable for $C$.*

Lemma 3.6 is easy to prove in the special case where all players are equidistant from the single facility. The case where different players can have different distances from the facility does not easily reduce to the uniform case, since changing the distances to be "more uniform" can decrease the summability of PT cost shares. For this reason, we instead prove Lemma 3.6 directly via a charging argument.

## 3.2  An $O(\log^2 k)$-Approximate Mechanism for Single-Sink Rent-or-Buy

Next we consider single-sink rent-or-buy (SSRoB) cost-sharing problems. The input is given by a graph $G = (V, E)$ with edge costs that satisfy the Triangle Inequality, a root vertex $t$, a set $U$ of demands (the players), each of which is located at a vertex of $G$, and a parameter $M \geq 1$. A feasible solution to the SSRoB problem induced by $S$ is a way of installing sufficient capacity on the edges of $G$ so that every player in $S$ can simultaneously route one unit of flow to $t$. Installing $x$ units of capacity on an edge $e$ costs $c_e \cdot \min\{x, M\}$; the parameter $M$ can be interpreted as the ratio between the cost of "buying" infinite capacity for a flat fee and the cost of "renting" a single unit of capacity. The cost $C(S)$ of a subset $S \subseteq U$ of players is then defined as the cost of an optimal solution to the SSRoB problem induced by $S$.

When $M = 1$, the SSRoB problem is equivalent to the Steiner tree problem. Theorem 4.2 in Section 4 thus implies that every $O(1)$-budget-balanced Moulin mechanism for $k$-player SSRoB cost-sharing problems is $\Omega(\log^2 k)$-approximate. Our next main result is a matching upper bound for the Moulin mechanism based on a cost-sharing method ("GST cost shares") discovered independently by Gupta, Srinivasan, and Tardos [9] and Leonardi and Schäfer [17].

**Theorem 3.7** *Let $C$ be a $k$-player SSRoB cost function and $\chi_{GST}$ the corresponding GST cost-sharing method. Then $\chi_{GST}$ is $O(\log^2 k)$-summable for $C$.*

Since GST cost shares are $O(1)$-budget-balanced and cross-monotonic [9, 17], Theorems 2.4 and 3.7 implies that the corresponding Moulin mechanism is $O(\log^2 k)$-approximate.

**Remark 3.8**  GST cost shares are defined as expectations of quantities that arise in a randomized algorithm. Unfortunately, it is not clear how to compute these cost shares in polynomial time. Gupta, Srinivasan, and Tardos [9] derandomized this algorithm and showed how to compute $O(1)$-budget balanced, cross-monotonic cost shares for the SSRoB problem in polynomial time. We expect that Theorem 3.7 will also hold for these cost shares, but have not yet verified this.

**Remark 3.9**  In this section we focused on facility location and SSRoB cost functions, but our techniques also apply to additional problems such as connected facility location [17] and edge cover [10]. Among the combinatorial optimization problems known to admit cross-monotonic cost-sharing methods with reasonable budget-balance, approximately efficient Moulin mechanisms are now known for all of them except for the Steiner forest problem [15].

## 4  An $\Omega(\log^2 k)$ Lower Bound for the Steiner Tree Problem

An instance of the *Steiner tree cost-sharing problem* [11] is given by an undirected graph $G = (V, E)$ with a root vertex $t$ and nonnegative edge costs, with each player of $U$ located at some vertex of $G$. For a subset $S \subseteq U$, the cost $C(S)$ is defined as that of a minimum-cost subgraph of $G$ that spans all of the players of $S$ as well as the root $t$. The following theorem is established in [23].

**Theorem 4.1 ([23])** *The JV Steiner tree mechanism is 2-budget-balanced and $O(\log^2 k)$-approximate.*

A matching lower bound for the JV mechanism was also shown in [23]. The main result of this section is a matching lower bound for *every* Moulin mechanism.

**Theorem 4.2** *There is a constant $c > 0$ such that the following statement holds. For every $\beta \geq 1$, every $\beta$-budget-balanced Moulin mechanism $M$ for Steiner tree cost-sharing problems, and every $k \geq 2$, there is a Steiner tree cost function $C$ defined on a universe $U$ and a subset $S \subseteq U$ such that $|S| = k$ and the output of $M$ has social cost at least $(c \log^2 k/\beta) \cdot C(S)$.*

In short, $O(1)$-budget-balanced Moulin mechanisms for Steiner tree cost-sharing problems must be $\Omega(\log^2 k)$-approximate. This lower bound also applies to the more general SSRoB cost-sharing problems, which proves that the Moulin mechanism of Subsection 3.2 achieves the minimum-possible worst-case efficiency loss (up to a constant factor). In light of Corollary 3.4 and the proof in [23] that the Shapley mechanism is $\mathcal{H}_k$-approximate for submodular cost-sharing problems, Theorem 4.2 implies that Steiner tree and SSRoB cost-sharing problems are fundamentally more difficult for Moulin mechanisms than facility location and submodular cost-sharing problems.

We now outline the proof of Theorem 4.2. At the highest level, our goal is to exhibit a (large) network $G$ such that every $O(1)$-budget-balanced Steiner tree Moulin mechanism behaves like the JV mechanism on some subnetwork of $G$.

Fix values for the parameters $k \geq 2$ and $\beta \geq 1$. We construct a sequence of networks, culminating in $G$. The network $G_0$ consists of a set $V_0$ of two nodes, one of which is the root $t$, which are connected by an edge of cost 1. The player set $U_0$ is $\sqrt{k}$ players that are co-located at the non-root node. (Assume for simplicity that $k$ is a power of 4.) For $i > 0$, we obtain the network $G_j$ from $G_{j-1}$ by replacing each edge $(v, w)$ of $G_{j-1}$ with $m$ internally disjoint two-hop paths between $v$ and $w$, where $m$ is a sufficiently large function of $k$ of $\beta$. (We will choose $m \geq 8\beta\sqrt{k} \cdot (2\beta)^{\sqrt{k}}$.) See Figure 1 in Appendix A. The cost of each of these $2m$ edges is half of the cost of the edge $(v, w)$. Thus every edge in $G_j$ has cost $2^{-j}$.

Let $V_j$ denote the vertices of $G_j$ that are not also present in $G_{j-1}$. We augment the universe by placing $\sqrt{k}$ new co-located players at each vertex of $V_j$; denote these new players by $U_j$. The final network $G$ is then $G_p$, where $p = (\log k)/2$. Let $V = V_0 \cup \cdots \cup V_p$ and $U = U_0 \cup \cdots \cup U_p$ denote the corresponding vertex and player sets. Let $C$ denote the corresponding Steiner tree cost function.

Now fix $\beta \geq 1$ and an arbitrary cross-monotonic, $\beta$-budget balanced Steiner tree cost-sharing method $\chi$. By Theorem 2.4, we can prove Theorem 4.2 by exhibiting a subset $S \subseteq U$ of size $k$ and an ordering $\sigma$ of the players of $S$ such that $\sum_{\ell=1}^{k} \chi(i_\ell, S_\ell) \geq (c \log^2 k/\beta) \cdot C(S)$, where $i_\ell$ and $S_\ell$ denote the $\ell$th player and the first $\ell$ players with respect to $\sigma$.

We construct the set $S$ iteratively. For $j = 0, 1, \ldots, p$, we will identify a subset $S_j \subseteq U_j$ of players; the set $S$ will then be $S_0 \cup \cdots \cup S_p$. Recall that $U_j$ consists of groups of $\sqrt{k}$ players, each co-located at a vertex of $V_j$, with $m$ such groups for each edge of $G_{j-1}$. The set $S_j$ will consist of zero or one such group of $\sqrt{k}$ players for each edge of $G_{j-1}$.

The set $S_0$ is defined to be $U_0$. For $j > 0$, suppose that we have already defined $S_0, \ldots, S_{j-1}$. Call a vertex $v \in V_0 \cup \cdots \cup V_{j-1}$ *active* if $v$ is the root $t$ or if the $\sqrt{k}$ players co-located at $v$ were included in the set $S_0 \cup \cdots \cup S_{j-1}$. Call an edge $(v, w)$ of $G_{j-1}$ *active* if both of its endpoints are active and *inactive* otherwise.

To define $S_j$, we consider each edge $(v, w)$ of $G_{j-1}$ in an arbitrary order. Each such edge gives rise to $m$ groups of $\sqrt{k}$ co-located players in $G_j$. If $(v, w)$ in inactive in $G_{j-1}$, then none of these $m\sqrt{k}$ players are included in $S_j$. If $(v, w)$ is active in $G_{j-1}$, then we will choose precisely one of the $m$ groups of players, and will include these $\sqrt{k}$ co-located players in $S_j$. We first state two lemmas that hold independently of how this choice is made; we then elaborate on our criteria for choosing groups of players.

**Lemma 4.3** *For every $j \in 1, 2, \ldots, p$, $|S_j| = 2^{j-1}\sqrt{k}$. Also, $|S_0| = \sqrt{k}$.*

Lemma 4.3 implies that $|S| = \sqrt{k}(1+\sum_{j=0}^{p-1} 2^j) = k$. The next lemma states that our construction maintains the invariant that the players selected in the first $j$ iterations lie "on a straight line" in $G$.

**Lemma 4.4** *For every $j \in 0, 1, \ldots, p$, $C(S_0 \cup \cdots \cup S_j) = 1$.*

In particular, Lemma 4.4 implies that $C(S) = 1$. Lemmas 4.3 and 4.4 both follow from straightforward inductions on $j$.

We now explain how to choose one out of the $m$ groups of co-located players that arise from an active edge. Fix an iteration $j > 0$ and let $\hat{S}$ denote the set of players selected in previous iterations $(S_0, \ldots, S_{j-1})$ and previously in the current iteration. Let $(v, w)$ be the active edge of $G_{j-1}$ under consideration and $A_1, \ldots, A_m \subseteq U_j$ the corresponding groups of co-located players. We call the group $A_r$ *good* if the $\sqrt{k}$ players of $A_r$ can be ordered $i_1, i_2, \ldots, i_{\sqrt{k}}$ so that

$$\chi(i_\ell, \hat{S} \cup \{i_1, \ldots, i_\ell\}) \geq \frac{1}{4\beta} \cdot \frac{2^{-j}}{\ell} \tag{2}$$

for every $\ell \in \{1, 2, \ldots, \sqrt{k}\}$. We then include an arbitrary good group $A_r$ in the set $S_j$.

The success of this approach crucially depends on the following lemma.

**Lemma 4.5** *Provided $m$ is a sufficiently large function of $k$ and $\beta$, for every $j \in \{1, \ldots, p\}$, every ordering of the active edges of $G_{j-1}$, and every edge $(v, w)$ in this ordering, at least one of the $m$ groups of players of $U_j$ that corresponds to $(v, w)$ is good. Also, the group $S_0$ is good.*

We prove Lemma 4.5 in Appendix C. We conclude by using the lemma to finish the proof of Theorem 4.2.

We have already defined the subset $S \subseteq U$ of players. We define the ordering $\sigma$ of the players in $S$ as follows. First, for all $j \in \{1, \ldots, p\}$, all players of $S_{j-1}$ precede all players of $S_j$ in $\sigma$. Second, for each $j \in \{1, \ldots, p\}$, the players of $S_j$ are ordered according to groups, with the $\sqrt{k}$ players of a group appearing consecutively in $\sigma$. The ordering of the different groups of players of $S_j$ is the same as the corresponding ordering of the active edges of $G_{j-1}$ that was used to define these groups. Third, each group of $\sqrt{k}$ co-located players is ordered so that (2) holds. This is possible by the definition of a good group and because all groups of players within $S$ are good (Lemma 4.5).

Now consider the sum $\sum_{\ell=1}^{k} \chi(i_\ell, S_\ell)$, where $i_\ell$ and $S_\ell$ denote the $\ell$th player and the first $\ell$ players of $S$ with respect to $\sigma$, respectively. Since (2) holds for every group of players, for every $j \in \{0, 1, \ldots, p\}$, every group of players in $S_j$ contributes at least

$$\sum_{\ell=1}^{\sqrt{k}} \frac{1}{4\beta} \cdot \frac{2^{-j}}{\ell} = \frac{2^{-j} \mathcal{H}_{\sqrt{k}}}{4\beta}$$

to this sum. By Lemma 4.3, for each $j \in \{1, \ldots, p\}$, there are $2^{j-1}$ such groups. There is also the group $S_0$. Thus the sum $\sum_{\ell=1}^{k} \chi(i_\ell, S_\ell)$ is at least

$$\frac{\mathcal{H}_{\sqrt{k}}}{4\beta} \left(1 + \sum_{j=1}^{(\log k)/2} 2^{j-1} \cdot 2^{-j}\right) \geq \frac{c}{\beta} \log^2 k = \left(\frac{c}{\beta} \log^2 k\right) \cdot C(S)$$

for some constant $c > 0$ that is independent of $k$ and $\beta$. This completes the proof of Theorem 4.2.

# 5  Beyond Groupstrategyproof Mechanisms

In this section we show that relaxing the constraint of groupstrategyproofness permits fundamentally better simultaneous approximation of budget-balance and efficiency. We begin with a particularly dramatic example: set cover cost-sharing problems. While every Moulin mechanism has extremely poor (inverse polynomial) budget-balance and efficiency, we show that a strategyproof mechanism due to Devanur, Mihail, and Vazirani [2] achieves a logarithmic approximation of both objective functions. We also show that a second strategyproof mechanism in [2] is both 1.861-budget-balanced and $O(\log k)$-approximate; this matches the efficiency bound of the Pál-Tardos mechanism (Theorem 3.3) while the lower bound of 3 that applies to all Moulin mechanisms for the problem [10].

**Efficiency of the DMV Set-Cover Mechanism.**  In this subsection we consider Set Cover cost-sharing function. Such a function is defined implicitly by a universe $U$ of players and a collection $\mathcal{C}$ of subsets of $U$, each of which has a given nonnegative cost. The cost $C(S)$ of a subset $S \subseteq U$ is then defined as the cost of minimum-cost subcollection of $\mathcal{C}$ whose union covers all players of $S$.

Immorlica, Mahdian, and Mirrokni [10] showed that every Moulin mechanism for Set Cover problems is $\Omega(\sqrt{k})$-budget-balanced. By Theorem 2.4, every such mechanism is also $\Omega(\sqrt{k})$-approximate. We next consider the strategyproof mechanism proposed by Devanur, Mihail and Vazirani [2] for Set Cover cost-sharing problems (the *DMV mechanism*), and show that it achieves a far superior (logarithmic) approximation of both objectives.

We now briefly review the DMV mechanism, which is based on the classical greedy algorithm for Set Cover and bears some resemblance to Moulin mechanisms. Initially, bids are collected from the players. After initializing a set $S$ to be $U$, the following steps are repeated until all of the players of $S$ have been marked. (All players are initially unmarked.) Every iteration, we choose a set $S_j \in \mathcal{C}$ that minimizes the ratio of the cost $c_j$ the set and the number $m_j$ of unmarked players of $S$ that it covers. We then try to charge each unmarked player $S_j \cap S$ a common cost share of $c_j/m_j \cdot \mathcal{H}_k$. If the bids of all unmarked players in $S_j \cap S$ are at least this amount, then all of these players are marked and the set $S_j$ is included in our final set cover solution. Otherwise, all of these players are deleted from the set $S$.

Devanur, Mihail, and Vazirani [2] proved the following: For every $k$-player set cover cost-sharing problem, the DMV mechanism is strategyproof and $\mathcal{H}_k$-budget balanced. In addition, the DMV mechanism assigns cost shares that lie in the core, and is *weakly groupstrategyproof* in the following sense: no coalition can bid in way that strictly improves the utility of all of the deviating players [2].

We show that in addition to satisfying these properties, the mechanism is $O(\log k)$-approximate.

**Theorem 5.1** *For every $k$-player Set Cover cost-sharing problem, the DMV mechanism is $(\mathcal{H}_k+1)$-approximate.*

The theorem follows immediately from two lemmas, each of which bounds a term of the social cost of the allocation constructed by DMV. The first lemma shows that the allocation returned by the DMV set cover mechanism is bounded above by $\mathcal{H}_k$ times the social cost of the optimal allocation. The proof of this lemma follows from three facts. First, the cost shares of the selected players pay for at least a $1/\mathcal{H}_k$ fraction of the cost of the constructed solution. Second, the cost shares assigned by the DMV mechanism are in the core. Third, players who are allocated service by the DMV mechanism pay at most their valuations. The second lemma upper bounds the total valuation of the players that are deleted by the DMV mechanism but included in the optimal solution. In the iteration during which such a player is deleted, the cost shares offered to it must exceed its valuation. By leveraging the greedy nature of the mechanism, these cost shares can then be charged to the cost of the optimal allocation.

Theorem 5.1 also applies to the special case of Vertex Cover cost-sharing problems; even for this special case, every Moulin mechanism is $\Omega(k^{1/3})$-budget-balanced [10] and therefore (by Theorem 2.4) $\Omega(k^{1/3})$-approximate.

**Efficiency of the DMV Facility Location Mechanism.** In this subsection, we give a second example of how relaxing the GSP constraint permits mechanisms with better approximation factors. Theorem 3.3 shows that the Moulin mechanism based on the Pál-Tardos cost-sharing scheme [21] is 3-budget-balanced and $O(\log k)$-approximate for facility location cost-sharing problems. Immorlica, Mahdian, and Mirrokni [10] showed that no such Moulin mechanism has better budget-balance, no matter how poor its approximate efficiency. We next show that the weakly groupstrategyproof mechanism proposed by Devanur, Mihail, and Vazirani [2] for facility location problems has better budget-balance and comparable approximate efficiency.

As in their Set Cover mechanism, the DMV mechanism for facility location cost-sharing problems is based on a greedy algorithm. We defer the details of the mechanism to Appendix D.2. Devanur, Mihail, and Vazirani [2] proved the following: For every $k$-player facility location cost-sharing problem, the DMV mechanism is weakly groupstrategyproof and 1.861-budget balanced. In addition, the DMV mechanism assigns cost shares that lie in the core.

We prove that the DMV mechanism is also $O(\log k)$-approximate.

**Theorem 5.2** *For every $k$-player facility location cost-sharing problem, the DMV mechanism is $(H_k/1.861 + 1.861)$-approximate.*

# References

[1] A. Archer, J. Feigenbaum, A. Krishnamurthy, R. Sami, and S. Shenker. Approximation and collusion in multicast cost sharing. *Games and Economic Behavior*, 47(1):36–71, 2004.

[2] N. R. Devanur, M. Mihail, and V. V. Vazirani. Strategyproof cost-sharing mechanisms for set cover and facility location games. In *Proceedings of the Fourth ACM Conference on Electronic Commerce (EC)*, pages 108–114, 2003.

[3] J. Feigenbaum, A. Krishnamurthy, R. Sami, and S. Shenker. Hardness results for multicast cost sharing. *Theoretical Computer Science*, 304:215–236, 2003.

[4] J. Feigenbaum, C. Papadimitriou, and S. Shenker. Sharing the cost of multicast transmissions. *Journal of Computer and System Sciences*, 63(1):21–41, 2001. Preliminary version in *STOC '00*.

[5] J. Green, E. Kohlberg, and J. J. Laffont. Partial equilibrium approach to the free rider problem. *Journal of Public Economics*, 6:375–394, 1976.

[6] A. Gupta, A. Kumar, M. Pál, and T. Roughgarden. Approximation via cost-sharing: A simple approximation algorithm for the multicommodity rent-or-buy problem. In *Proceedings of the 44th Annual Symposium on Foundations of Computer Science (FOCS)*, pages 606–615, 2003.

[7] A. Gupta, A. Kumar, and T. Roughgarden. Simpler and better approximation algorithms for network design. In *Proceedings of the 35th Annual ACM Symposium on the Theory of Computing (STOC)*, 2003.

[8] A. Gupta, M. Pál, R. Ravi, and A. Sinha. Boosted sampling: Approximation algorithms for stochastic optimization. In *Proceedings of the 36th Annual ACM Symposium on the Theory of Computing (STOC)*, pages 417–426, 2004.

[9] A. Gupta, A. Srinivasan, and É. Tardos. Cost-sharing mechanisms for network design. In *Proceedings of the 7th International Workshop on Approximation Algorithms for Combinatorial Optimization Problems (APPROX)*, volume 3122 of *Lecture Notes in Computer Science*, pages 139–150, 2004.

[10] N. Immorlica, M. Mahdian, and V. S. Mirrokni. Limitations of cross-monotonic cost-sharing schemes. In *Proceedings of the 16th Annual ACM-SIAM Symposium on Discrete Algorithms (SODA)*, pages 602–611, 2005.

[11] K. Jain and V. Vazirani. Applications of approximation algorithms to cooperative games. In *Proceedings of the 33rd Annual ACM Symposium on the Theory of Computing (STOC)*, pages 364–372, 2001.

[12] K. Jain and V. Vazirani. Equitable cost allocations via primal-dual-type algorithms. In *Proceedings of the 34th Annual ACM Symposium on the Theory of Computing (STOC)*, pages 313–321, 2002.

[13] D. R. Karger and M. Minkoff. Building Steiner trees with incomplete global knowledge. In *Proceedings of the 41st Annual Symposium on Foundations of Computer Science (FOCS)*, pages 613–623, 2000.

[14] K. Kent and D. Skorin-Kapov. Population monotonic cost allocation on mst's. In *Operational Research Proceedings KOI*, pages 43–48, 1996.

[15] J. Könemann, S. Leonardi, and G. Schäfer. A group-strategyproof mechanism for Steiner forests. In *Proceedings of the 16th Annual ACM-SIAM Symposium on Discrete Algorithms (SODA)*, pages 612–619, 2005.

[16] J. Könemann, S. Leonardi, G. Schäfer, and S. van Zwam. From primal-dual to cost shares and back: A stronger LP relaxation for the steiner forest problem. In *Proceedings of the 32nd Annual International Colloquium on Automata, Languages, and Programming (ICALP)*, volume 3580 of *Lecture Notes in Computer Science*, pages 1051–1063, 2005.

[17] S. Leonardi and G. Schäfer. Cross-monotonic cost-sharing methods for connected facility location. In *Proceedings of the Fifth ACM Conference on Electronic Commerce (EC)*, pages 242–243, 2004.

[18] A. Mas-Colell, M. D. Whinston, and J. R. Green. *Microeconomic Theory*. Oxford University Press, 1995.

[19] H. Moulin. Incremental cost sharing: Characterization by coalition strategy-proofness. *Social Choice and Welfare*, 16:279–320, 1999.

[20] H. Moulin and S. Shenker. Strategyproof sharing of submodular costs: Budget balance versus efficiency. *Economic Theory*, 18:511–533, 2001.

[21] M. Pál and É. Tardos. Group strategyproof mechanisms via primal-dual algorithms. In *Proceedings of the 44th Annual Symposium on Foundations of Computer Science (FOCS)*, pages 584–593, 2003.

[22] K. Roberts. The characterization of implementable choice rules. In J. J. Laffont, editor, *Aggregation and Revelation of Preferences*. North-Holland, 1979.

[23] T. Roughgarden and M. Sundararajan. New trade-offs in cost-sharing mechanisms. In *Proceedings of the 38th Annual ACM Symposium on the Theory of Computing (STOC)*, 2006. To appear.

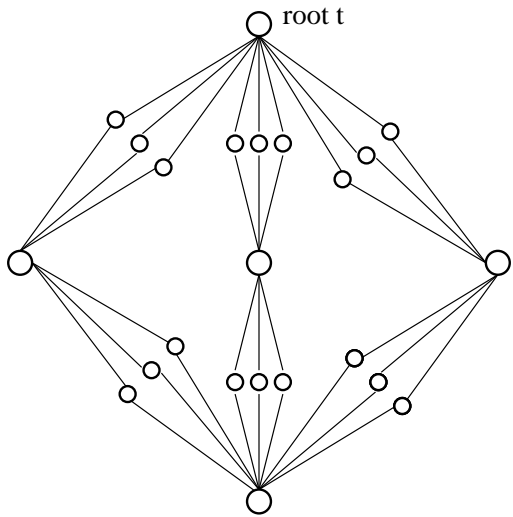

Figure 1: The figure shows an intermediate stage in the graph construction process for $m = 3$. This value of $m$ is chosen only for convenience of depiction. The above graph, $G_2$, is obtained after two iterations of the construction process. All the edges have length $\frac{1}{4}$.

## A   Figures

## B   Missing Proofs from Section 3

### B.1   Missing Proofs from Subsection 3.1

#### B.1.1   Proof of Proposition 3.1

*Proof of Proposition 3.1:* Fix $k \geq 1$. Let $C$ denote the cost function corresponding to a facility location instance in which $k$ players are co-located with a facility that has opening cost 1. Thus $C(\emptyset) = 0$ and $C(S) = 1$ if $S \neq \emptyset$. Let $\chi$ be a $\beta$-budget-balanced, cross-monotone cost-sharing method for $C$.

Let $\sigma$ denote a random permutation of the players. We can complete the proof by showing that the expected value of the left-hand side of (1) is at least $\mathcal{H}_k/\beta$. For $\ell \in \{1, \ldots, k\}$, let $X_\ell$ denote the random variable equal to $\chi(i_\ell, S_\ell)$, where $S_\ell$ and $i_\ell$ denote the set of the first $\ell$ players and the $\ell$th player with respect to $\sigma$, respectively. Fix $\ell$ and condition on the set $S_\ell$. Since $\chi$ is $\beta$-budget-balanced and every player of $S_\ell$ is equally likely to be last, the conditional expectation of $X_\ell$ is at least $1/\beta\ell$. Since this inequality holds independently of the value of $S_\ell$, $\mathbf{E}\left[X_\ell\right] \geq 1/\beta\ell$. Summing over $\ell$ and applying linearity of expectation completes the proof. ∎

#### B.1.2   Proof of Lemma 3.5

Before proving Lemma 3.5, we review the Pál-Tardos cost-sharing methods for facility location problems [21]. Given a facility location instance defined by players $U$, facilities $F$ with opening costs $f$, and a metric $c$ on $U \cup F$, the corresponding PT cost-sharing method $\chi_{PT}$ is defined as follows. Fix an arbitrary subset $S \subseteq U$ of players. First, there is a notion of *time*, which is initially 0 and increases at a uniform rate. At a time $t \geq 0$, we associate with each player $i \in S$ a ball of radius $t$ with center $j$, where distances are with respect to the given metric $c$. Once a ball includes a facility $q \in F$, the subsequent growth of this ball contributes toward "filling" this facility. Once these contributions equal the facility's opening cost $f_q$, we declare the facility $q$ to be *full*. Precisely, facility $q$ becomes full at the time $t^q$ defined by the equation

$$\sum_{i \in S} \max\{0, t^q - c(q,i)\} = f_q. \tag{3}$$

The PT cost share $\chi_{PT}(i, S)$ of a player $i$ in $S$ is then defined as the length of time during which there is no full facility in player $i$'s ball. These cost-shares and known to be both 3-budget-balanced and cross-monotone [21].

Next, we establish that increasing distances can only increase PT cost shares.

**Lemma B.1** *Let $\mathcal{I}$ and $\mathcal{I}'$ denote two instances of uncapacitated facility location with the same player set $U$, facility sets $F$, and facility opening costs $f$. Assume that the metric $c'$ on $U \cup F$ of the second instance dominates the first, in that $c'(i,j) \geq c(i,j)$ for every $i, j \in U \cup F$. Let $\chi_{PT}$ and $\chi'_{PT}$ be the PT cost-sharing methods corresponding to $\mathcal{I}$ and $\mathcal{I}'$, respectively. Then $\chi'_{PT}(i, S) \geq \chi_{PT}(i, S)$ for every set $S \subseteq U$ and player $i \in S$.*

*Proof:* Fix a set $S \subseteq U$. First, equation (3) immediately implies facilities can only become full later in the instance $\mathcal{I}'$ than in $\mathcal{I}$. Second, note that the PT cost share of a player $i \in S$ is defined as the earliest time at which a full facility lies in player $i$'s ball. It follows that PT cost shares for $\mathcal{I}'$ can only be larger than those for $\mathcal{I}$. ∎

*Proof of Lemma 3.5:* Fix an arbitrary facility location instance $\mathcal{I}^0$, given by a player set $U$, a facility set $F$, facility opening costs $f$, and a metric $c^0$ on $U \cup F$. Let $C^0$ denote the corresponding facility location cost function, $\chi^0_{PT}$ the corresponding PT cost-sharing method, and $\alpha$ the smallest number such that $\chi^0_{PT}$ is $\alpha$-summable for $C^0$. We aim to exhibit a single-facility instance such that the corresponding PT cost-sharing method is no better than $\alpha$-summable.

Choose an ordering $\sigma$ of the players of $U$ and a set $S \subseteq U$ such that $\sum_{\ell=1}^{|S|} \chi^0_{PT}(i_\ell, S_\ell) = \alpha \cdot C^0(S)$, where $S_\ell$ and $i_\ell$ denote the set of the first $\ell$ players and the $\ell$th player of $S$ in the ordering $\sigma$, respectively.

Fix an optimal solution to the facility location instance induced by the players in $S$. Let $F^*$ denote the facilities opened by this solution, and let $S_q$ denote the players of $S$ assigned to the facility $q \in F^*$ in this solution. This solution has cost

$$C^0(S) = \sum_{q \in F^*} \left( f_q + \sum_{i \in S_q} c^0(i, q) \right).$$

We obtain the instance $\mathcal{I}^1$ from $\mathcal{I}^0$ by modifying distances as follows. If a player $i \in S$ is assigned to the facility $q \in F^*$ in the fixed optimal solution to the instance induced by $S$, then set $c^1(i, q) = c^0(i, q)$. All other distances between players and facilities are set to a sufficiently large number. Other distances, which will not play a role in what follows, can be set as large as possible subject to the Triangle Inequality. Let $C^1$ denote the cost function corresponding to $\mathcal{I}^1$ and $\chi^1_{PT}$ the corresponding PT cost-sharing method. By construction, $C^1(S) = C^0(S)$. Lemma B.1 implies that

$$\sum_{\ell=1}^{|S|} \chi^1_{PT}(i_\ell, S_\ell) \geq \alpha \cdot C^1(S), \tag{4}$$

where $S_\ell$ and $i_\ell$ denote the first $\ell$ players and the $\ell$th player of $S$ with respect to $\sigma$. Thus $\chi^1_{PT}$ is no better than $\alpha$-summable for $C^1$.

The instance $\mathcal{I}^1$ is essentially a collection of independent single-facility instances. To make this precise, for a facility $q \in F^*$, let $\mathcal{I}^q$ denote the facility location instance with player set $S^q$, facility set $\{q\}$, opening cost $f_q$, and with distances inherited from $\mathcal{I}_1$. Let $C^q$ and $\chi^q_{PT}$ denote the corresponding cost function and PT cost-sharing method, respectively. By construction, we have

$$C^1(S) = \sum_{q \in F^*} C^q(S^q). \tag{5}$$

Further, our definition of the distances in $\mathcal{I}^1$ ensures that the PT cost share $\chi^1(i_\ell, S_\ell)$ for a player $i_\ell \in S^q$ is a function only of the set of other players of $S^q$ that precede $i_\ell$ in the ordering $\sigma$. (Only players of $S^q$ contribute to the filling of facility $q$.) Because of this, we have

$$\sum_{\ell=1}^{|S|} \chi^1_{PT}(i_\ell, S_\ell) = \sum_{q \in F^*} \sum_{p=1}^{|S^q|} \chi^q_{PT}(i_p, S^q_p), \tag{6}$$

where $S^q_p$ and $i_p$ denote the first $p$ players and the $p$th player of $S_q$, respectively, according to $\sigma$. Inequality (4) and equations (5) and (6) imply that for some $q \in F^*$,

$$\sum_{p=1}^{|S^q|} \chi^q_{PT}(i_p, S^q_p) \geq \alpha \cdot C^q(S^q).$$

The PT cost shares $\chi^q$ are therefore no better than $\alpha$-summable for the single-facility instance $\mathcal{I}^q$, and the proof is complete. ∎

### B.1.3 Proof of Lemma 3.6

*Proof of Lemma 3.6:* Consider an arbitrary facility location problem with a single facility $q$ and a set $U$ of $k$ players. Let $f_q$ denote the opening cost for $q$ and $c(j, q)$ denote the distance between the player $j$ and the facility $q$. For a set $S$ of players, let $c(S, q)$ denote the sum of the distances of the demands in the set $S$ to the facility $q$: $c(S, q) = \sum_{i \in S} c(i, q)$. Because there is only one facility, we have $C(S) = f_q + c(S, q)$ for every $S \subseteq U$. Let $\chi_{PT}$ denote the PT cost-sharing method.

Fix an arbitrary ordering $\sigma$ of the players in $U$ and a subset $S \subseteq U$. Let $S_\ell$ be the first $\ell$ players of $S$ in the ordering and $i_\ell$ the $\ell$th player. By Definition 2.3, we need to show that

$$\sum_{\ell=1}^{|S|} \chi_{PT}(i_\ell, S_\ell) \leq H_k \cdot C(S). \tag{7}$$

Fix $\ell \in \{1, 2, \ldots, |S|\}$ and consider the set $S_\ell$ of players. Recall from the definition of PT cost shares that there is a time $t$ at which the facility $q$ becomes full. There are two cases. If $c(i_\ell, q) > t$, then the PT cost share $\chi(i_\ell, S_\ell)$ equals $c(i_\ell, q)$—by the time $q$ lies in player $i_\ell$'s ball, it is already full. If $c(i_\ell, q) \leq t$, then the PT cost share $\chi_{PT}(i_\ell, S_\ell)$ is $t$—by the time facility $q$ is full, it already lies in player $i_\ell$'s ball.

In the latter case, the growth of player $i_\ell$'s ball contributes toward filling the facility $q$ during the time interval $[c(i_\ell, q), t]$. Since $q$ is the only facility and it is not full during this time, the cost shares of all of the players in $S_\ell$ are accumulating during this time. All but $c(i, q)$ of the increase in the cost share of a player $i \in S_{\ell-1}$ during this time must contribute toward the filling of facility $q$. Thus,

$$[t - c(i_\ell, q)] + \sum_{i \in S_{\ell-1}} [t - c(i_\ell, q) - c(i, q)] \leq f_q.$$

Rewriting,

$$t - c(i_\ell, q) \leq \frac{1}{\ell} \left( f_q + \sum_{i \in S_{\ell-1}} c(i, q) \right).$$

We can therefore bound the PT cost share $\chi_{PT}(i_\ell, S_\ell)$ of player $i_\ell$ by

$$\chi_{PT}(i_\ell, S_\ell) \leq c(i_\ell, q) + \frac{1}{\ell}\left(f_q + \sum_{i \in S_{\ell-1}} c(i, q)\right).$$

Summing over all $\ell \in \{1, 2, \ldots, |S|\}$ then gives

$$\begin{aligned}
\sum_{\ell=1}^{|S|} \chi_{PT}(i_\ell, S_\ell) &\leq \sum_{\ell=1}^{|S|}\left[c(i_\ell, q) + \frac{1}{\ell}\left(f_q + \sum_{i \in S_{\ell-1}} c(i, q)\right)\right] \\
&= f_q \sum_{\ell=1}^{|S|} \frac{1}{\ell} + \sum_{\ell=1}^{|S|} c(i_\ell, q) \cdot \left(1 + \sum_{p=\ell+1}^{|S|} \frac{1}{p}\right) \\
&\leq \mathcal{H}_k \cdot \left(f_q + \sum_{\ell=1}^{|S|} c(i_\ell, q)\right) \\
&= \mathcal{H}_k \cdot C(S),
\end{aligned}$$

completing the proof of (7). ∎

## B.2 Proof of Theorem 3.7

To prepare for the proof of Theorem 3.7, we recapitulate the GST cost-sharing method of [9, 17]. Fix an SSRoB cost function $C$, defined implicitly via the graph $G = (V, E)$ with edge costs $c$ and root vertex $t$, the parameter $M \geq 1$, and the player set $U$. Fix a set $S \subseteq U$ and choose a random subset $D \subseteq S$ be adding each player $i \in S$ to $D$ independently with probability $1/M$. Condition on the set $D$. The conditional cost share of a player $i \notin D$ is defined as the shortest-path distance between $i$ and a player in $D \cup \{t\}$. The conditional cost share of a player $i \in D$ is defined as $M$ times the Jain-Vazirani cost share $\chi_{JV}(i, D)$ of $i$ with respect to the Steiner tree instance defined by $G$, $c$, $t$, and the players $D$ (see [11] and Section 4 for details). The GST cost share $\chi_{GST}(i, S)$ of player $i \in S$ is then defined as the expected value of its conditional cost shares, where the expectation is over the random choice of the subset $D$. These cost shares are known to be 4-budget-balanced and cross-monotonic [9, 17].

Next, note that a GST cost share $\chi_{GST}(i, S)$ can be naturally decomposed into the sum of two terms: a term $\chi_{buy}(i, S)$ that corresponds to choices of the subset $D$ that include $i$, and a term $\chi_{rent}(i, S)$ corresponding to subsets $D$ that exclude $i$. Mirroring several recent analyses of sampling algorithms for rent-or-buy problems [7, 6, 8], our analysis proceeds in two steps. First, we directly upper bound the summability of $\chi_{buy}$. Second, show that the summability of $\chi_{rent}$ is at most a constant factor times that of $\chi_{buy}$.

We first show that for every SSRoB cost function, the corresponding cost-sharing method $\chi_{buy}$ is $O(\log^2 k)$-summable.

**Lemma B.2** *Let $C$ be a $k$-player SSRoB cost function and $\chi_{buy}$ the first term of the corresponding GST cost-sharing method. Then $\chi_{buy}$ is $O(\log^2 k)$-summable for $C$.*

*Proof:* Fix an ordering $\sigma$ of $U$ and a set $S \subseteq U$ of players. Condition on the random choice of the set $D \subseteq S$. Since JV cost shares are $O(\log^2 k)$-summable for all Steiner tree cost functions [23], the sum of the conditional cost shares of players in $D$ is $O(M \log^2 k)$ times the cost of an optimal Steiner tree spanning $D \cup \{t\}$:

$$\sum_{\ell=1}^{|D|} \chi_{JV}(i_\ell, D_\ell) \leq O(\log^2 k) \cdot M \cdot OPT_D,$$

where $OPT_D$ denotes the cost of an optimal Steiner tree spanning $D \cup \{t\}$, and $D_\ell$ and $i_\ell$ denote the set of the first $\ell$ players and the $\ell$th player, respectively, of the set $D$ according to $\sigma$.

A result of Gupta, Kumar, and Roughgarden [7, Lemma 2.2], based on earlier work by Karger and Minkoff [13], implies that $M$ times the expectation (over $D$) of $OPT_D$ is at most the cost $C(S)$ of an optimal SSRoB solution. Thus

$$\mathbf{E}_D \left[ \sum_{\ell=1}^{|D|} \chi_{JV}(i_\ell, D_\ell) \right] \leq O(\log^2 k) \cdot C(S). \tag{8}$$

For a player $i \in S$, let $X_i$ denote the random variable equal to 0 when $i \notin D$, and equal to the term $\chi_{JV}(i_\ell, D_\ell)$ that corresponds to player $i$ when $i \in D$. Note that $\mathbf{E}_D[X_i] = \chi_{buy}(i, S_i)$, where $S_i$ is the set of players of $S$ equal to or preceding player $i$ in the ordering $\sigma$. Applying linearity of expectation to the left-hand side of (8) then completes the proof. ∎

Next we show that the summability of the second term $\chi_{rent}$ of the GST cost-sharing method is at most a constant factor times that of the first. To prove this, we make use of the following simple fact about the JV Steiner tree cost-sharing method.

**Fact B.3** *Let $C$ be the Steiner tree cost function defined by the graph $G = (V, E)$ with edge costs $c$, root vertex $t$, and player set $U$. Let $\chi_{JV}$ be the corresponding JV cost-sharing method. Then for every subset $S \subseteq U$ and player $i \in S$, the JV cost share $\chi_{JV}(i, S)$ of $i$ is at least half of the shortest-path distance between $i$ and some other player in $S \cup \{t\}$.*

We now bound the summability of the $\chi_{rent}$ term of the GST cost-sharing method, which will complete the proof of Theorem 3.7.

**Lemma B.4** *Let $C$ be a $k$-player SSRoB cost function and $\chi_{rent}$ the first term of the corresponding GST cost-sharing method. Then $\chi_{rent}$ is $O(\log^2 k)$-summable for $C$.*

*Proof:* Fix an ordering $\sigma$ of $U$, a set $S \subseteq U$ of players, and a player $i_\ell \in S$ that is $\ell$th in the ordering $\sigma$. Let $S_\ell$ denote the players of $S$ that equal or precede $i$ in the ordering $\sigma$. Recall that, after conditioning on the random subset $D \subseteq S_\ell$ of players, player $i_\ell$'s conditional cost share contributes to $\chi_{buy}(i_\ell, S_\ell)$ if $i_\ell \in D$ and to $\chi_{rent}(i_\ell, S_\ell)$ if $i_\ell \notin D$.

For every player $i$ preceding $i_\ell$ in the ordering $\sigma$, condition on whether or not $i$ is included in the random sample, and let $D' \subseteq S_\ell \setminus \{i_\ell\}$ be the selected players. Player $i_\ell$ will be included in the set $D$ with probability $1/M$, in which case its conditional cost share will be $M \cdot \chi_{JV}(i_\ell, D' \cup \{i_\ell\})$, where $\chi_{JV}(i_\ell, D' \cup \{i_\ell\})$ is player $i_\ell$'s JV cost share in the Steiner tree cost-sharing problem induced by the players of $D' \cup \{i_\ell\}$. Player $i_\ell$ is excluded from the random sample $D$ with probability $(1-1/M)$, in which case its conditional cost share equals the shortest distance $d(i_\ell, D' \cup \{t\})$ between $i_\ell$ and either a player of $D'$ or the root vertex $t$.

Let $R_\ell$ denote the random variable equal to $d(i_\ell, D \cup \{t\})$ if $i_\ell \notin D$ and equal to 0 otherwise. Let $B_\ell$ denote the random variable equal to $M \cdot \chi_{JV}(i_\ell, D)$ if $i_\ell \in D$ and equal to 0 otherwise. We then have

$$\begin{aligned}
\chi_{rent}(i_\ell, S_\ell) &= \mathbf{E}_D[R_\ell] \\
&= \mathbf{Pr}[i \notin D] \cdot \mathbf{E}_{D'}\left[d(i_\ell, D' \cup \{t\})\right] \\
&\leq 2 \cdot \mathbf{E}_{D'}\left[\chi_{JV}(i_\ell, D' \cup \{i_\ell\}\right] \\
&= 2 \cdot \mathbf{Pr}[i \in D] \cdot \mathbf{E}_{D'}\left[M \cdot \chi_{JV}(i_\ell, D' \cup \{i_\ell\}\right] \\
&= 2 \cdot \mathbf{E}_D[B_\ell] \\
&= 2 \cdot \chi_{buy}(i_\ell, S_\ell),
\end{aligned}$$

where the inequality follows from Fact B.3. Summing over all players $i_\ell \in S$ and applying Lemma B.2 completes the proof. ∎

## C  Missing Proofs from Section 4

*Proof of Lemma 4.5:* Fix an iteration $j \in \{1, \ldots, p\}$, an ordering of the active edges of $G_{j-1}$, and an edge $(v, w)$ in this ordering. Let $A_1, \ldots, A_m$ denote the $m$ groups of $\sqrt{k}$ co-located players of $U_j$ corresponding to the edge $(v, w)$. Let $\hat{S}$ denote the players already included in $S$ in previous and the present iteration. Let $X_1$ denote the union $A_1 \cup \cdots \cup A_m$.

We claim that for $m$ sufficiently large, at least $m/2\beta$ of the groups $A_1, \ldots, A_m$ satisfy

$$\sum_{i \in A_r} \chi(i, \hat{S} \cup X_1) \geq \frac{2^{-j}}{4\beta}. \tag{9}$$

We say that such groups *survive*. Every surviving group must contain a player $i$ for which $\chi(i, \hat{S} \cup X_1) \geq 2^{-j}/4\beta\sqrt{k}$.

To prove the claim, label the surviving groups $A_1^1, \ldots, A_q^1$. Let $Y_1$ denote their union $A_1^1 \cup \cdots \cup A_q^1$. By Lemma 4.4, we have $C(\hat{S}) = 1$—in words, there is a subgraph $H$ of $G$ that spans all of the players of $\hat{S}$ and that has total cost 1. Since $(v, w)$ is active, the vertices $v$ and $w$ are both spanned by $H$. Since each group $A_1^1, \ldots, A_q^1$ is connected to both $v$ and $w$ by edges of cost $2^{-j}$, $C(\hat{S} \cup Y_1) \leq 1 + q2^{-j}$. Indeed, the structure of $G$ ensures that this inequality holds with equality. We then have

$$\begin{aligned}\sum_{i \in \hat{S} \cup Y_1} \chi(i, \hat{S} \cup X_1) &\leq \sum_{i \in \hat{S} \cup Y_1} \chi(i, \hat{S} \cup Y_1) \\ &\leq C(\hat{S} \cup Y_1) \\ &= 1 + q2^{-j},\end{aligned} \tag{10}$$

where the first and second inequalities follows from the cross-monotonicity and approximate budget-balance of $\chi$, respectively.

Since (9) fails for non-surviving groups, and there at most $m$ such groups, we have

$$\sum_{i \in X_1 \setminus Y_1} \chi(i, \hat{S} \cup X_1) \leq \frac{m 2^{-j}}{4\beta}. \tag{11}$$

Combining (10) and (11) then gives

$$\sum_{i \in \hat{S} \cup X_1} \chi(i, \hat{S} \cup X_1) \leq 1 + 2^{-j}\left(q + \frac{m}{4\beta}\right). \tag{12}$$

On the other hand, $C(\hat{S} \cup X_1) = 1 + m2^{-j}$ as noted above. Since $\chi$ is $\beta$-budget-balanced, we have

$$\sum_{i \in \hat{S} \cup X_1} \chi(i, \hat{S} \cup X_1) \geq \frac{1}{\beta}\left(1 + m2^{-j}\right). \tag{13}$$

Combining (12) and (13) and rearranging gives the constraint

$$q \geq \frac{3}{4\beta}m - 2^j\left(1 - \frac{1}{\beta}\right) \geq \frac{3}{4\beta}m - \sqrt{k}\left(1 - \frac{1}{\beta}\right).$$

Thus $q \geq m/2\beta$ provided $m$ is a sufficiently large function of $k$ and $\beta$, as claimed.

We have thus identified $q \geq m/2\beta$ surviving groups $A_1^1, \ldots, A_q^2$, each of which contains a player $i$ for which $\chi(i, \hat{S} \cup X_1) \geq 2^{-j}/4\beta\sqrt{k}$. We next repeat this process. More precisely, for each $r \in \{1, \ldots, q\}$, obtain $A_r^2$ from the surviving group $A_r^1$ by removing an arbitrary player $i$ with $\chi(i, \hat{S} \cup X_1) \geq 2^{-j}/4\beta\sqrt{k}$. Let $X_2$ denote the union $A_1^2 \cup \cdots \cup A_q^2$. The above argument implies that, as long as $q \geq m/2\beta$ is a sufficiently large function of $k$ and $\beta$, then at least $q/2\beta$ of the sets $A_1^2, \ldots, A_q^2$ survive by satisfying an analogue of inequality (9), with the sets $X_1$ and $A_r$ replaced by $X_2$ and $A_r^2$, respectively.

Choose $m \geq 8\beta\sqrt{k} \cdot (2\beta)^{\sqrt{k}}$. Iterating this procedure and reindexing the surviving groups after each iteration, we inductively obtain a collection of disjoint sets $A_1^h, \ldots, A_{q_h}^h$ for each $h \in \{1, 2, \ldots, \sqrt{k}\}$ with the following properties:

(1) $q_h \geq m/(2\beta)^h$;

(2) for each $r \in \{1, \ldots, q_h\}$, $A_r^h$ contains a player $i_r^h$ such that $\chi(i_r^h, \hat{S} \cup X_h) \geq 2^{-j}/4\beta(\sqrt{k} - h + 1)$, where $X_h = \cup_r A_r^h$;

(3) for each $r \in \{1, \ldots, q_h\}$ and $h > 1$, $A_r^h = A_r^{h-1} \setminus \{i^{h-1}\}$.

By (1) and our choice of $m$, $q_{\sqrt{k}} \geq 1$. By properties (2) and (3) and cross-monotonicity of $\chi$, the group $A_1^1$ that corresponds to $A_1^{\sqrt{k}}$ is good in the sense of (2). The proof is complete. ∎

# D  Missing Proofs from Section 5

## D.1  Proof of Theorem 5.1

Fix an instance of the Set-Cover Game $I \equiv \{U, T, c\}$. Fix a vector **v** of valuations. Let $S$ denote the set serviced by DMV when players bid truthfully. Let $C'(S)$ denote the cost that DMV incurs in servicing $S$. Let $O^*$ be the set of players which minimizes social cost.

We need to show that:

$$C'(S) + V(U \setminus S) \leq (H_k + 1)(C(O^*) + V(U \setminus O^*))$$

We start by bounding $C'(S)$:

**Lemma D.1** $C'(S) \leq H_k \cdot (C(O^*) + V(S \setminus O^*))$

*Proof:* First note that $\frac{1}{H_k}C'(S)$ is completely paid for by the members of $S$.

$$\sum_{i \in S} \chi_{DMV}(i, S) \geq \frac{1}{H_k}C'(S)$$

$$\sum_{i \in O^* \cap S} \chi_{DMV}(i, S) + \sum_{i \in S \setminus O^*} \chi_{DMV}(i, S) \geq \frac{1}{H_k}C'(S).$$

Since the cost-shares are in the core the first term on the left hand side of the above inequality can be bounded above by $C(O^* \cap S)$, which is at most $C(O^* \cap S)$ as the cost function $C$ is increasing. Since players bid truthfully and DMV satisfies VP and NPT, the second term on the left hand side of the above inequality can be bounded by $V(S \setminus O^*)$. The lemma follows. ∎

Next we bound $V(O^* \setminus S)$:

**Lemma D.2** $V(O^* \setminus S) \leq C(O^*)$

*Proof:* Let $\sigma$ denote the order in which players in $O^* \setminus S$ were deleted by DMV. Let $l$ denote the size of $O^* \setminus S$. When player $\sigma_i$ is deleted by DMV, it is offered a price at most $\frac{C(O^*)}{(l-i+1)H_k}$. Also, since players are bidding truthfully the valuation $v_{\sigma_i}$ is less than the price offered at deletion. Summing over all the players in $O^* \setminus S$ gives $V(O^* \setminus S) \leq \frac{H_l}{H_k}C(O^*)$. Since $k \geq l$, we have the lemma. ∎

The theorem follows easily from the lemmas.

## D.2 Proof of Theorem 5.2

Fix an instance of the facility location problem $I \equiv \{F, D, c\}$. We start by giving a brief description of the DMV facility location mechanism. First, the mechanism accepts bids. The description of the mechanism involves the notion of time. At time $t = 0$, all cost shares are initialized to 0. The mechanism then raises the cost shares at a uniform rate. At any time instant $t$ in the process, a demand which is unconnected and still under consideration pays an amount $max(0, t - c_{ij})$ toward the opening cost of any unopened facility $j$. During the course of execution three types of events arise. First, the cost share of a player may exceed its bid. In this case the player is deleted and removed from further consideration. Second, cost shares contributed toward an unopened facility may equal its opening cost. In this case, the facility is declared open and all unconnected demands which make non zero contributions toward the opening cost are connected to the facility. The demands are charged their current cost share. Third, the cost share of an unconnected demand may equal its connection cost to some opened facility. In this case, the demand is connected to the facility and is charged its current cost share. In the latter cases, when a demand is connected to a facility, its contributions toward the opening cost of other facilities is withdrawn. Likewise, when a demand is deleted, all of its contributions toward facility opening costs are withdrawn. The mechanism terminates when all the demands are either serviced or deleted. To ensure that the cost shares are in the core, all the costs (connection and facility opening costs) are initially scaled by a factor $1.861$. We are now ready to prove Theorem 5.2.

Fix a vector **v** of valuations. Let $S$ denote the set serviced by DMV when players bid truthfully. Let $C'(S)$ denote the cost that DMV incurs in servicing $S$. Let $O^*$ be the set of players which minimizes social cost.

We need to show that:

$$C'(S) + V(U \setminus S) \leq (\frac{H_k}{1.861} + 1.861)(C(O^*) + V(U \setminus O^*))$$

We start by bounding $C'(S)$:

**Lemma D.3** $C'(S) \leq 1.861 \cdot (C(O^*) + V(S \setminus O^*))$

*Proof:* First note that $\frac{1}{1.861}C'(S)$ is completely paid for by the members of $S$.

$$\sum_{i \in S} \chi_{DMV}(i, S) \geq \frac{1}{1.861}C'(S)$$

$$\sum_{i \in O^* \cap S} \chi_{DMV}(i, S) + \sum_{i \in S \setminus O^*} \chi_{DMV}(i, S) \geq \frac{1}{1.861}C'(S).$$

Since the cost-shares are in the core the first summand on the left hand side of the above inequality can be bounded above by $C(O^* \cap S)$, which is at most $C(O^* \cap S)$ as the cost function $C$ is increasing. Since players bid truthfully and DMV satisfies VP and NPT, the second summand on the left hand side of the above inequality can be bounded by $V(S \setminus O^*)$. The lemma follows. ∎

Next we bound $V(O^* \setminus S)$. Let $\sigma$ denote the order in which players in $O^* \setminus S$ were deleted by DMV. Let $\sigma_i$ denote the $i^{th}$ player in the ordering. Let $x_{DMV}(j)$ denote the price offered by DMV when the player $j$ was deleted.

Let $O$ an allocation that optimizes social cost. The optimal solution $O$ consists of a set of stars, each with a facility and many demands connected to it. Each demand belongs to a specific star. For any player $\sigma_i$ deleted by DMV, let $O_{\sigma_i}$ denote the set of players from the same star in the optimal solution, which belong to $O^* \setminus S$) and are after $\sigma_i$ in the ordering $\sigma$. By the correctness of the DMV algorithm we have that:

**Fact D.4** *The valuations of players in $O_{\sigma_i} \cup \{\sigma_i\}$ must be at least $x_{DMV}(\sigma_i) - \epsilon$ for any $\epsilon > 0$.*

Let $f_{\sigma_i}$ denote the facility in the star to which $\sigma_i$ belongs. Consider a run of greedy algorithm on the set of demands $O_{\sigma_i} \cup \{\sigma_i\}$ with facility $f_{\sigma_i}$, with the opening costs and the metric being scaled down by a factor 1.861. Let $x_{OPT}(\sigma_i)$ denote the time at which the player $\sigma_i$ is served. Then:

**Lemma D.5** $x_{DMV}(\sigma_i) \le x_{OPT}(\sigma_i)$

*Proof:* Proof sketch. Assume this is not true: $x_{DMV}(\sigma_i) > x_{OPT}(\sigma_i)$.

In the run on the star, by time $x_{OPT}(\sigma_i)$, the facility $f_{\sigma_i}$ must be full *and* the demand $\sigma_i$ should touch it. Since all the players in $O_{\sigma_i} \cup \{\sigma_i\}$ are also present in the run of DMV at time $x_{DMV}(\sigma_i)$, they were also present at time $x_{OPT}(\sigma_i)$. So, at this time, the facility $f_{\sigma_i}$ would have filled up and $\sigma_i$ would touch it. By Fact D.4, it should have been serviced. This is a contradiction. ∎

**Fact D.6** *On single facility instances, the cost-shares offered by the Pal-Tardos are identical to the those offered by the greedy algorithm.*

**Lemma D.7** $V(O^* \setminus S) \le \frac{H_k}{1.861} C(O^*)$

*Proof:* Since players are bidding truthfully the valuation $v_{\sigma_i} < x_{DMV}(\sigma_i)$. By Lemma D.5, $v_{\sigma_i} < x_{OPT}(\sigma_i)$. By Fact D.6 and Lemma 3.6, summing over all the players in $O^* \setminus S$ gives $V(O^* \setminus S) \le \frac{H_k}{1.861} \cdot C(O^*)$. ∎

The theorem follows easily from the lemmas.